\documentclass[
  preprint,
  superscriptaddress,
  amsmath,amssymb,aps,showpacs,
  twoside,final,secnumarabic,
  nofootinbib
]{revtex4-2}
\pdftrailerid{<4C476154722D444C4350323032363030>}

\usepackage[
  paperwidth=205mm,
  paperheight=290mm,
  top=17mm,
  bottom=25mm,
  inner=17mm,
  outer=17mm,
  twoside
]{geometry}

\usepackage[T1]{fontenc}
\usepackage[utf8]{inputenc}
\usepackage[english]{babel}
\usepackage{graphicx}
\usepackage{array}[=2016-10-06]
\usepackage{dcolumn}
\usepackage{bm}
\usepackage{floatpag}
\usepackage{float}
\usepackage{multirow}
\usepackage{placeins}
\usepackage{url}
\usepackage{fancyhdr}
\usepackage[
  unicode=true,
  colorlinks=true,
  linkcolor=magenta,
  urlcolor=blue,
  citecolor=blue,
  breaklinks=true
]{hyperref}

\DeclareGraphicsExtensions{.pdf,.png,.eps}

\newcommand{\DLCPVolume}{\textbf{80}}
\newcommand{\DLCPIssue}{9}
\newcommand{\DLCPReportNumber}{17}
\floatpagestyle{fancy}

\begin{document}

\newcommand{\DLCPPlaceholder}[1]{\texttt{[DLCP placeholder: #1]}}
\newcommand{\DLCPFCNCMetricsGate}[1]{\texttt{[FCNC metrics gate: #1]}}

\newcommand{\DLCPConferenceTitle}{Lorentz-Equivariant Geometric Algebra
  Transformers for Top-Quark Event Reconstruction}
\title{Machine Learning in Fundamental Physics\\[20pt]
\DLCPConferenceTitle}
\hypersetup{
  pdftitle={\DLCPConferenceTitle},
  pdfauthor={E. Abasov, L. Dudko, E. Iudin, A. Markina, M. Perfilov,
    P. Volkov, G. Vorotnikov, A. Zaborenko}
}

\newcommand{\DLCPAffiliation}{Federal State Budget Educational Institution of Higher Education M.V.Lomonosov Moscow State University, Skobeltsyn Institute of Nuclear Physics (SINP MSU), 1(2), Leninskie gory, GSP-1, Moscow 119991, Russian Federation}

\author{\firstname{E.}~\surname{Abasov}}
\email[E-mail: ]{emil@abasov.ru}
\affiliation{\DLCPAffiliation}
\author{\firstname{L.}~\surname{Dudko}}
\affiliation{\DLCPAffiliation}
\author{\firstname{A.}~\surname{Markina}}
\affiliation{\DLCPAffiliation}
\author{\firstname{M.}~\surname{Perfilov}}
\affiliation{\DLCPAffiliation}
\author{\firstname{P.}~\surname{Volkov}}
\affiliation{\DLCPAffiliation}
\author{\firstname{G.}~\surname{Vorotnikov}}
\affiliation{\DLCPAffiliation}
\author{\firstname{A.}~\surname{Zaborenko}}
\affiliation{\DLCPAffiliation}

\received{}
\revised{}
\accepted{}

\begin{abstract}
Results are presented for Lorentz-equivariant geometric algebra transformers
(LGaTr) in three event reconstruction tasks involving top quarks. Geometric algebra provides a common Lorentz-covariant representation of four-momenta and typed physical objects. Performance of LGaTr is evaluated on three different physics tasks - $t\bar{t}/tWb$ separation, FCNC detection in single top-quark production and universal event reconstruction. LGaTr is outperforming the baseline networks in $tW/t\bar{t}$ separation and full event reconstruction and is showing comparable or better performance to standard transformer-based networks. The implementation is provided in a companion repository~\url{https://github.com/emil2001/top-quark-universal-nets}.
\end{abstract}

\par
\keywords{top quark; geometric algebra; transformers; event reconstruction; machine learning\\[5pt]}

\maketitle
\thispagestyle{fancy}

\providecommand{\DLCPBackMatter}{}
\section{Introduction}
\label{sec:introduction}

The reconstruction and interpretation of top quark events depend on relations
among four-momenta, particle identities, and decay chains.  Conventional
machine learning systems can infer these relations from coordinate components
and constructed observables.  Lorentz-equivariant representations instead
encode the transformation properties of event objects directly.  Geometric
algebra provides a common language for four-momenta and their covariant
combinations~\cite{1,2}.

The Lorentz-equivariant Geometric Algebra Transformer (LGaTr) is studied in
three complementary top quark event reconstruction tasks.  The first separates
single-resonant \(tW\) and double-resonant \(t\bar t\) contributions to a
common final state.  The second identifies flavour-changing neutral-current
single top quark production.  The third performs multiprocess reconstruction
from detector-level objects, jointly predicting parton-level particles, their
classes, their decay assignments, and high-level observables.  Together, these
tasks connect a controlled parton-level example, a realistic single top quark
classification problem, and a common reconstruction setting with multiple
outputs.

The paper is organised as follows.  Section~\ref{sec:geometric-algebra}
introduces the geometric-algebra representation and its physics motivation.
Section~\ref{sec:higen-example} describes the \(tW\)-versus-\(t\bar t\)
application.  Section~\ref{sec:fcnc} considers flavour-changing
neutral-current single top quark classification.  Section~\ref{sec:universal-model}
presents the universal event reconstruction task, and
Section~\ref{sec:conclusions} gives the conclusions.

\section{Geometric algebra for Lorentz-covariant event representation}
\label{sec:geometric-algebra}

\subsection{Algebraic language}

Geometric algebra provides a common algebraic language for Lorentz
covariance and collider-event kinematics.  We use the Clifford algebra
\(\mathrm{Cl}(1,3)\) of Minkowski spacetime with signature
\((+,-,-,-)\).  Its orthogonal grade-one basis \(\gamma_\mu\), where
\(\mu,\nu\in\{0,1,2,3\}\) and repeated raised and lowered spacetime
indices are summed, obeys
\[
  \gamma_\mu\gamma_\nu+\gamma_\nu\gamma_\mu=2\eta_{\mu\nu},
  \qquad
  \eta_{\mu\nu}=\mathrm{diag}(+1,-1,-1,-1).
\]
For two grade-one vectors \(p\) and \(q\), the geometric product is
\[
  pq=p\!\cdot\!q+p\wedge q,
  \qquad
  p\!\cdot\!q=\tfrac12(pq+qp),
  \qquad
  p\wedge q=\tfrac12(pq-qp),
\]
where \(p\!\cdot\!q\) is the Lorentz scalar inner product and
\(p\wedge q\) is the outer product.  For an on-shell \(W\) boson decay
into effectively massless leptons, \(2p_\ell\!\cdot p_\nu=m_W^2\), whereas
\(p_\ell\wedge p_\nu\) is a six-component bivector that represents the
oriented leptonic decay plane.

The algebra is graded as
\[
  \mathrm{Cl}(1,3)=\bigoplus_{k=0}^{4}
  \left\langle\mathrm{Cl}(1,3)\right\rangle_k,
  \qquad 1+4+6+4+1=16.
\]
Thus, scalar, vector, bivector, trivector, and pseudoscalar components
belong to a single multivector space.  Event-shape observables and other
auxiliary quantities can be supplied through scalar channels when required.

Proper, orthochronous Lorentz transformations act on a multivector \(X\)
through a rotor,
\[
  X\longmapsto R X\widetilde R,
  \qquad R=\exp(B/2),
  \qquad B\in\left\langle\mathrm{Cl}(1,3)\right\rangle_2,
  \qquad R\widetilde R=1,
\]
where the tilde denotes the Clifford reverse.  The rotor belongs to
\(\mathrm{Spin}^{+}(1,3)\), the spin cover of
\(\mathrm{SO}^{+}(1,3)\).  This sandwich action transforms every grade
consistently without first choosing a coordinate \(4\times4\) matrix
representation.

The closure of the algebra under geometric products motivates
Lorentz-equivariant network layers.  Equivariant bilinear maps on
multivector inputs can be constructed from geometric products followed by
grade projections.  The resulting model family is referred to as the
Lorentz-equivariant Geometric Algebra Transformer (LGaTr)~\cite{8}.

\subsection{Physics motivation and event encoding}

For a collider event with \(N\) typed objects, the input representation can
be written as
\[
  \mathcal E=\sum_{i=1}^{N}p_i\otimes\lvert f_i\rangle
  \in\mathrm{Cl}(1,3)\otimes V_{\mathrm{flav}},
\]
where \(p_i\) is the grade-one four-momentum of object \(i\),
\(\lvert f_i\rangle\) encodes its discrete object or flavour label, and
\(V_{\mathrm{flav}}\) is the corresponding label space.  The geometric
products of the four-momenta generate covariant components of higher grades,
while scalar auxiliary information can be included in scalar channels.

The scalar sector contains familiar invariant quantities, including
\(m_i^2=p_i\!\cdot p_i\) and
\(m_{ij}^2=(p_i+p_j)^2\).  With specified incoming and outgoing
four-momenta, the same inner products form the partonic Mandelstam scalars
\(\hat s\), \(\hat t\), and \(\hat u\).  Rotors also provide a common
notation for the boosts and rotations used to define helicity-frame angles
and spin-quantisation axes~\cite{5,9,10}.

The most representative example, where GA formalism simplifies calculations, are top quark spin correlations~\cite{5}. Standard approach involves: (1) boost 4-momenta to $t\bar{t}$ rest frame, (2) boost to respective $t/\bar{t}$ rest frame, (3) project lepton directions onto spin-quantization axis. In GA, this reduces to just one operation
$$
\cos\Delta\phi_{spin} \sim \frac{\langle B_t \widetilde{B_{\bar{t}}}\rangle_0}{|B_t||B_{\bar{t}}|}
$$
where $B_t = p_b \wedge p_l \wedge p_{\nu}$, $B_{\bar{t}} = p_{\bar{b}} \wedge p_{\bar{l}} \wedge p_{\bar{\nu}}$ and $\langle\rangle_0$ denotes grade projection. Hence, the assumption is that GA-based networks, such as LGaTr, can achieve better generalization and better performance for physics-based tasks with potentially lower parameter counts. This example illustrates how kinematic
vectors, decay planes, and scalar observables can be represented within one
Lorentz-covariant framework.

\section{\(tW\) and \(t\bar t\) discrimination}
\label{sec:higen-example}

\subsection{Task and representation}

The study conducted in~\cite{4} considers the parton-level separation of the
single-resonant \(tW\) and double-resonant \(t\bar t\) contributions to the
common \(pp\to tWb\) final state.  The two truth classes are defined at the
matrix-element level with a diagram-removal prescription.  The full
gauge-invariant \(tWb\) matrix-element sample, which contains both
contributions and their interference, is used as a cross-check rather than as
a training class.

The LGaTr classifier represents the six final-state particles with particle
tokens \(T_i\). Their grade-one component is
\(\langle T_i\rangle_1=p_i^\mu\gamma_\mu\), while grade-zero components encode
particle type, electric charge, and \(b\)-tag information.  Two
pairing-hypothesis tokens complement the particle tokens, giving eight tokens
in total.  Full details of the event generation, inputs, training, and model
architecture are given in Ref.~\cite{4}.

\subsection{Network comparison}

The compact LGaTr model used here is based on Ref.~\cite{8}. Its three
equivariant blocks combine a per-grade equivariant linear layer, a
geometric-product layer, and a Lorentz-invariant attention head. The
grade-zero coefficients are mean-pooled over tokens before a dense
classification head.

The reference is a fully connected multilayer perceptron (MLP) using about
75 engineered high-level collider features.  These combine true Lorentz
invariants, including scalar products and reconstructed masses, with
beam-frame or subgroup-invariant quantities, including transverse momentum
\(p_{\mathrm{T}}\), pseudorapidities, and angular observables.  Its dense
layout is \(75\to500\to500\to500\to1\), with rectified linear units and a
sigmoid output.  Table~\ref{tab:twtt-comparison} summarises the cited model
architectures~\cite{4}.

\begin{table}[t]
  \centering
  \caption{Architectures for the parton-level \(tW\) versus
    \(t\bar t\) example.  \(N_{\mathrm{par}}\) is the reported number of
    trainable parameters.}
  \label{tab:twtt-comparison}
  \begin{tabular}{lp{0.53\linewidth}r}
    \hline
    Model & Representation and architecture & \(N_{\mathrm{par}}\) \\
    \hline
    Reference MLP & \({\sim}75\) high-level features; \(75\to500\to500\to500\to1\)
      & \(5.4\times10^5\) \\
    LGaTr & six particle and two pairing-hypothesis tokens; three blocks
      & \(1.51\times10^5\) \\
    \hline
  \end{tabular}
\end{table}

\subsection{Results}

Table~\ref{tab:twtt-results} gives the area under the receiver operating
characteristic curve (AUC).  For the reference MLP, the reported uncertainty
is the Hanley--McNeil estimator for a single reference-MLP checkpoint, whereas
for LGaTr it is the standard deviation across LGaTr seeds 42--46.  These
uncertainties have different meanings, so the numerical comparison concerns
the central AUC values.  The LGaTr and reference-MLP central values are
\(0.9653\) and \(0.9594\), respectively, and
\(\Delta\mathrm{AUC}_{\mathrm{LGaTr-MLP}} = 0.0059\).

\begin{table}[t]
  \centering
  \caption{Area under the receiver operating characteristic curve for the
    parton-level \(tW\)-versus-\(t\bar t\) sample. }
  \label{tab:twtt-results}
  \begin{tabular}{lll}
    \hline
    Model & AUC \\
    \hline
    Reference MLP & \(0.9594 \pm 0.0005\) \\
    LGaTr & \(0.9653 \pm 0.0006\) \\
    \hline
  \end{tabular}
\end{table}

LGaTr uses fewer trainable parameters, and its central AUC is higher than that
of the reference MLP.

\section{Flavour-changing neutral-current single top quark application}
\label{sec:fcnc}

\subsection{Task and inputs}

We consider a flavour-changing neutral-current (FCNC) single top quark
classification task based on a CMS analysis~\cite{16}.
The signal arises through the \(tug\) FCNC coupling and is classified against
standard model background.  The background comprises \(t\)-, \(tW\)-, and
\(s\)-channel single top quark production, \(t\bar t\), \(W\)+jets, and quantum
chromodynamics (QCD) multijet processes. The QCD multijet component combines
simulated and data-driven events. Simulated samples are generated
with CompHEP~\cite{13}, POWHEG~\cite{Alioli:2010xd}, and
MadGraph5~\cite{Alwall:2011uj}; hadronisation is performed with
PYTHIA8~\cite{Bierlich:2022pfr}, followed by a detailed CMS detector
simulation~\cite{17}.

The low-level representation contains the ordered physical-object sequence
\(\ell,\nu,j_b,j_1,j_2\): a lepton, reconstructed neutrino, a b jet, and
up to two light jets.

The LGaTr model contains the five physical tokens, the fixed reference tokens
\(B_\perp=e_{12}\) and \(t=e_0\), and laboratory-frame event shapes.  For the
valid physical-object tokens with laboratory-frame spatial momenta
\(\vec p_k\), the normalised spatial momentum tensor is
\[
 \mathcal{S}_{\mathrm{mom}}^{ij}
 =\frac{\sum_k p_k^i p_k^j}{\sum_k \lvert\vec p_k\rvert^2},
 \qquad i,j\in\{x,y,z\},
 \qquad
 \lambda_1^{(\mathrm{mom})}\ge\lambda_2^{(\mathrm{mom})}
 \ge\lambda_3^{(\mathrm{mom})}.
\]
The resulting sphericity \(S\) and aplanarity \(A\) are
\[
 S=\frac{3}{2}\left(\lambda_2^{(\mathrm{mom})}
   +\lambda_3^{(\mathrm{mom})}\right),
 \qquad
 A=\frac{3}{2}\lambda_3^{(\mathrm{mom})}\,.
\]
These auxiliary grade-zero scalars provide additional event-level information
to the network.

The high-level MLP baseline instead uses a separate input of 54 variables,
including sphericity, planarity, and \(M_{tW}\), tailored for single top quark
samples.  The Transformer uses the five-token low-level representation without
additional kinematic features.

The geometric path retains equivariance under \(\mathrm{SO}(2)\) rotations
about the beam axis.  The fixed reference tokens encode the corresponding
detector reference directions.

\subsection{Models and screening with three training seeds}

\begin{figure}
    \centering
    \includegraphics[width=0.8\linewidth]{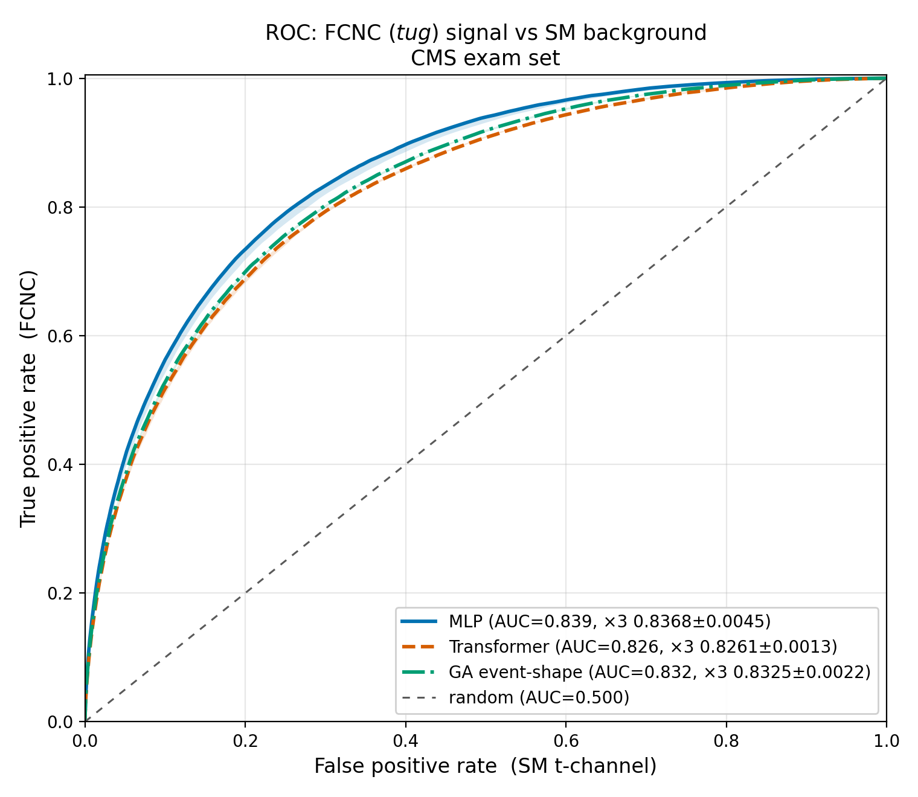}
    \caption{ROC curves for an FCNC sample with a mixed simulated and
    data-driven QCD multijet component. Solid, dashed, and dash-dotted curves
    denote the MLP, Transformer, and LGaTr, respectively; the diagonal denotes
    random classification. The MLP has the largest central score
    (Table~\ref{tab:fcnc-summary}).}
    \label{fig:fcnc-roc}
\end{figure}

Table~\ref{tab:fcnc-summary} reports FCNC scores using a ROC-AUC metric. Fig.~\ref{fig:fcnc-roc} shows the corresponding ROC curves. As can be seen from this comparison, LGaTr slightly outperforms standard transformer in this task, demonstrating the benefit of Lorentz-equivariant architecture on similar inputs. However, baseline MLP is still better overall, thus showing that the raw 4-momenta do not have the required amount of information about the event compared to hand-crafted features.

\begin{table}[t]
  \centering
  \small
  \caption{\footnotesize FCNC scores: ROC-AUC.}
  \label{tab:fcnc-summary}
  \begin{tabular}{l c}
    \hline
    Model & Three-seed mean \(\pm\) standard deviation \\
    \hline
    MLP & \(0.8368 \pm 0.0045\) \\
    LGaTr & \(0.8325 \pm 0.0022\) \\
    Transformer & \(0.8261 \pm 0.0013\) \\
    \hline
  \end{tabular}
\end{table}

\section{Universal event reconstruction model}
\label{sec:universal-model}

The universal event reconstruction task maps a detector-level input to
parton-level targets. The detector-level input contains missing transverse
momentum, up to four charged leptons and 12 jets; empty slots represent absent
objects.  Each object is represented by its four-momentum and scalar
identification information. The parton-level targets encode decays of top
quarks and \(W\) bosons. Regression reconstructs four-momenta for charged leptons, neutrinos, jets, top quarks, and \(W\) bosons. Classification determines particle identities, including empty slots. Assignment determines decay relations from each top quark to its \(W\) boson and b jet and from each \(W\) boson to its lepton, neutrino, or jet decay products. High-level regression (HLR) reconstructs derived event observables. In the FCNC comparison of Section~\ref{sec:fcnc}, these high-level variables are the principal source of the MLP advantage.

\begin{figure*}[t]
  \centering
  \includegraphics[width=0.95\textwidth]{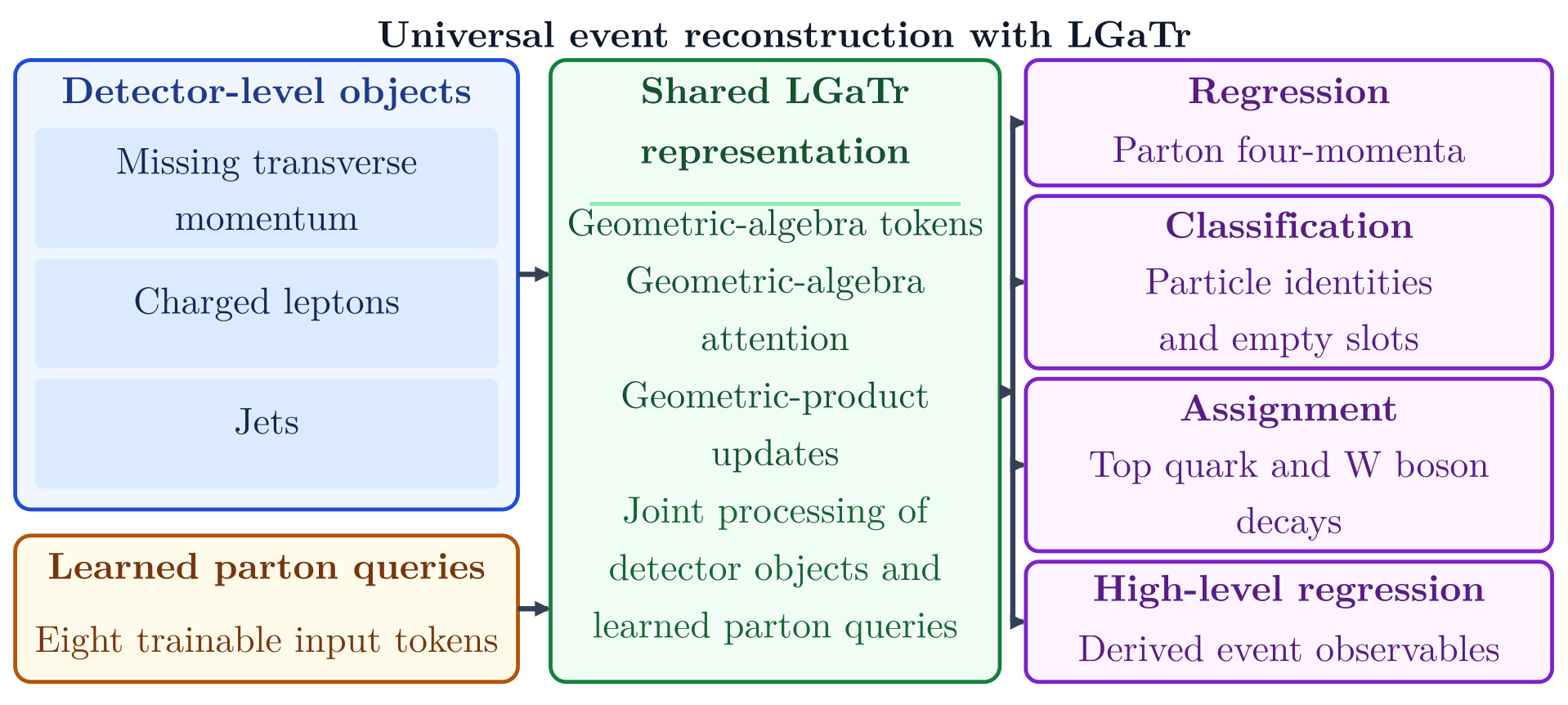}
\caption{Overview of the LGaTr universal event reconstruction model.
    Detector-level objects and eight learned parton queries are distinct inputs
    to the shared geometric-algebra processing.
    Separate readouts perform regression, classification, decay assignment, and
    high-level regression.}
  \label{fig:universal-overview}
\end{figure*}

Figure~\ref{fig:universal-overview} distinguishes detector-level objects from
learned parton queries and shows their shared geometric-algebra processing.

\subsection{Multiprocess task and architecture comparison}

The universal event reconstruction task uses a common detector-level input and
output across five sample categories: single top quark production in
the \(t\) channel, \(t\bar t\), events with three top quarks and a \(j\), events with
three top quarks and a \(W\) boson, and events with four top quarks.  Here, \(j\) denotes an additional jet.
The same detector-level input, parton-level targets, and four outputs are used
for every process category~\cite{18}.

\FloatBarrier
Table~\ref{tab:universal-configurations} compares the architectures along
their event representation and token-processing axes.  The multilayer
perceptron baseline flattens the event to 130 components.  The Detection
Transformer (DETR) instead encodes 17 detector tokens for missing transverse
momentum, leptons, and jets. It uses eight learned parton queries: four for top
quarks and four for \(W\) bosons.
Here a query denotes a learned token before readout, rather than a reconstructed
particle~\cite{19}.

\begin{table}[H]
  \centering
  \small
  \caption{Architectures for the common multiprocess reconstruction task.
    The entries are architecture and input descriptions, not performance
    results.}
  \label{tab:universal-configurations}
  \begin{tabular}{p{0.27\linewidth}p{0.63\linewidth}}
    \hline
    Model & Event representation and architecture \\
    \hline
    MLP & Flattened 130-component event; 4 dense blocks. \\
    DETR & 17 detector tokens and 8 learned queries for top quarks and \(W\) bosons;
      4 encoder layers, 4 decoder layers, and 4 attention heads. \\
    LGaTr & 17 detector multivector tokens, 8 learned parton queries,
      and 2 fixed reference tokens; 4 geometric-algebra attention blocks,
      each with 16 multivector channels and 4 attention heads. \\
    \hline
  \end{tabular}
\end{table}

In LGaTr, each detector token has six multivector channels: one grade-one
four-momentum channel and five grade-zero scalar channels: a jet b-tagging
score, lepton flavour, and binary type
indicators for missing transverse momentum, charged leptons, and jets.  Inputs
of grades 2 through 4 are initially empty
and are used only in the hidden layers.  The LGaTr stack processes the detector
tokens and learned parton queries together with two fixed reference tokens.
The eight learned parton query multivectors are unconstrained tensors: all 16
geometric-algebra components remain trainable.  They have a grade-zero
initialisation, but this does not constrain their subsequent values.  The
queries are fixed architectural inputs rather than values transformed with the
event.  This seed/query construction does not impose a transformation
constraint on the end-to-end reconstruction mapping.  The two fixed reference
tokens encode the transverse plane and time direction through the bivector
\(B_\perp=\gamma_1\gamma_2\) and time vector \(t=\gamma_0\), respectively.

\subsection{Results}

\begin{figure}
    \centering
    \includegraphics[width=0.95\linewidth]{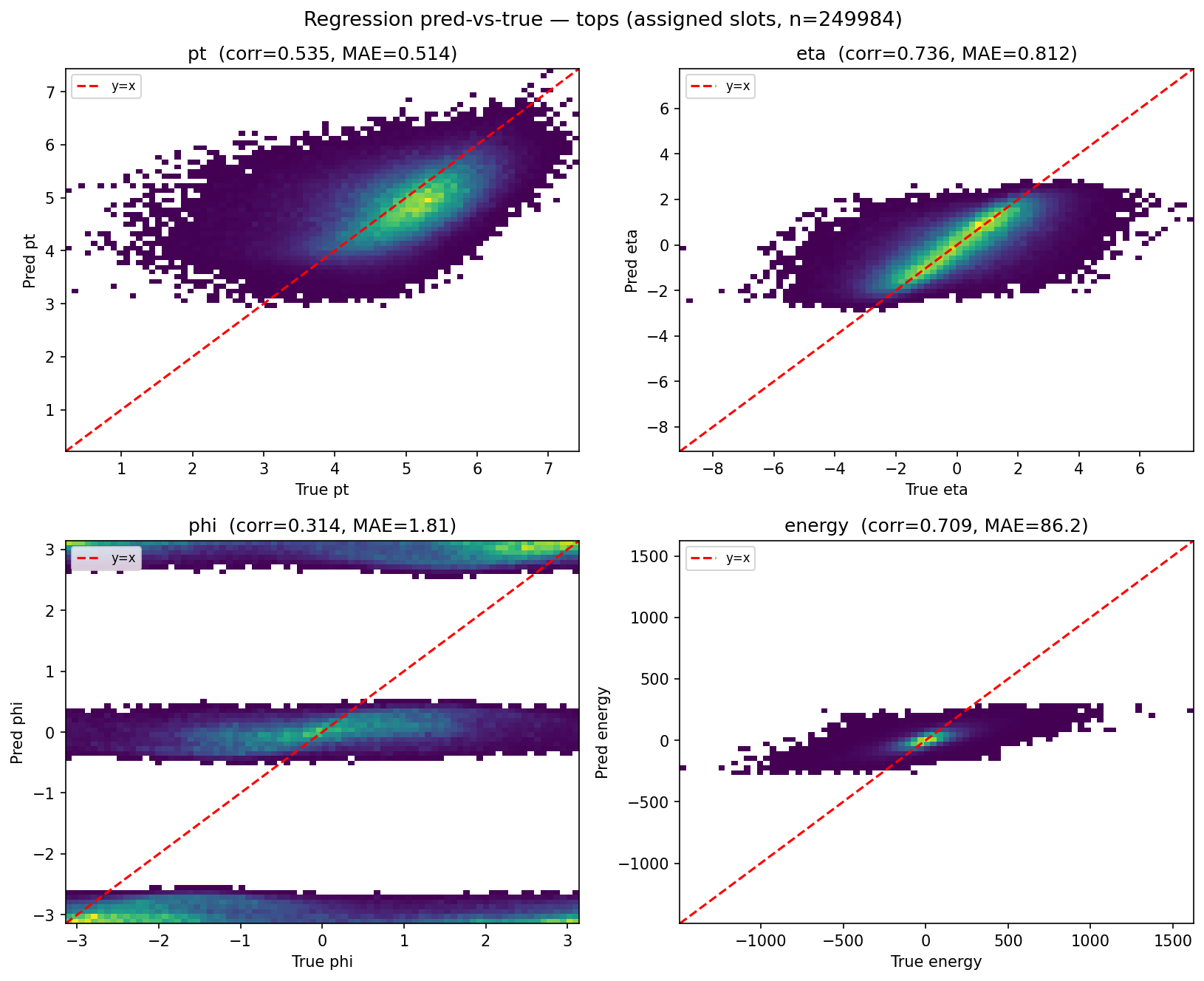}
    \caption{LGaTr reconstruction of the four-momentum of matched top quark
    targets (\(N=249\,984\)) in the universal event reconstruction sample.
    The panels show transverse momentum, pseudorapidity, azimuthal angle, and
    energy; transverse momentum and energy are converted from their
    standardized logarithmic target representation before plotting.  The
    horizontal and vertical axes give target and reconstructed physical values,
    respectively.  The dashed diagonal marks exact reconstruction.  The
    transverse-momentum, pseudorapidity, and energy panels list Pearson
    correlation coefficients and mean absolute deviations.  The azimuthal-angle
    panel lists the Fisher--Lee circular correlation (\(r_\mathrm{FL}\)) and mean
    absolute wrapped deviation.  Densities close to the diagonal, a larger
    correlation, and a smaller deviation indicate a better reconstruction.}
    \label{fig:top-reg}
\end{figure}

\begin{figure}[!htbp]
    \centering
    \setlength{\medskipamount}{0pt}
    \setlength{\abovecaptionskip}{0pt}
    \includegraphics[width=0.45\linewidth]{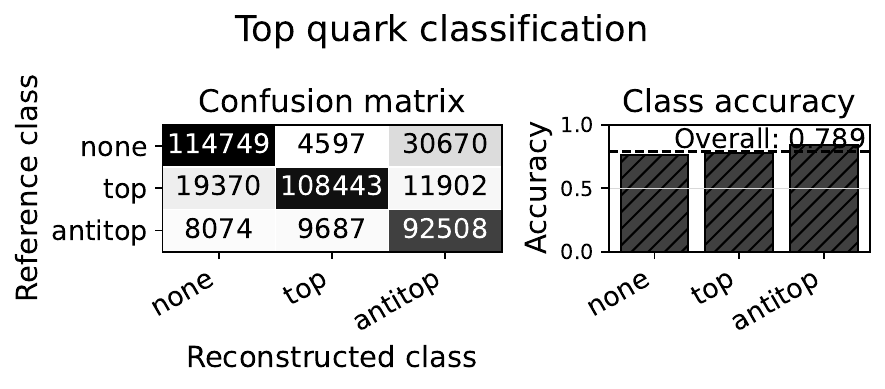}\par\medskip
    \includegraphics[width=0.95\linewidth]{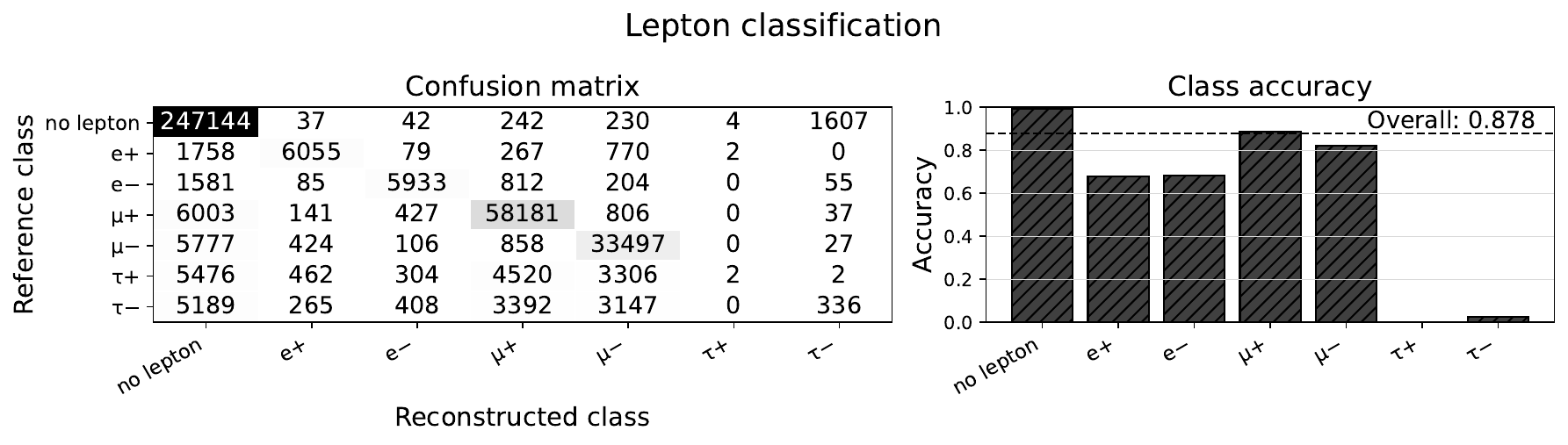}\par\medskip
    \includegraphics[width=0.55\linewidth]{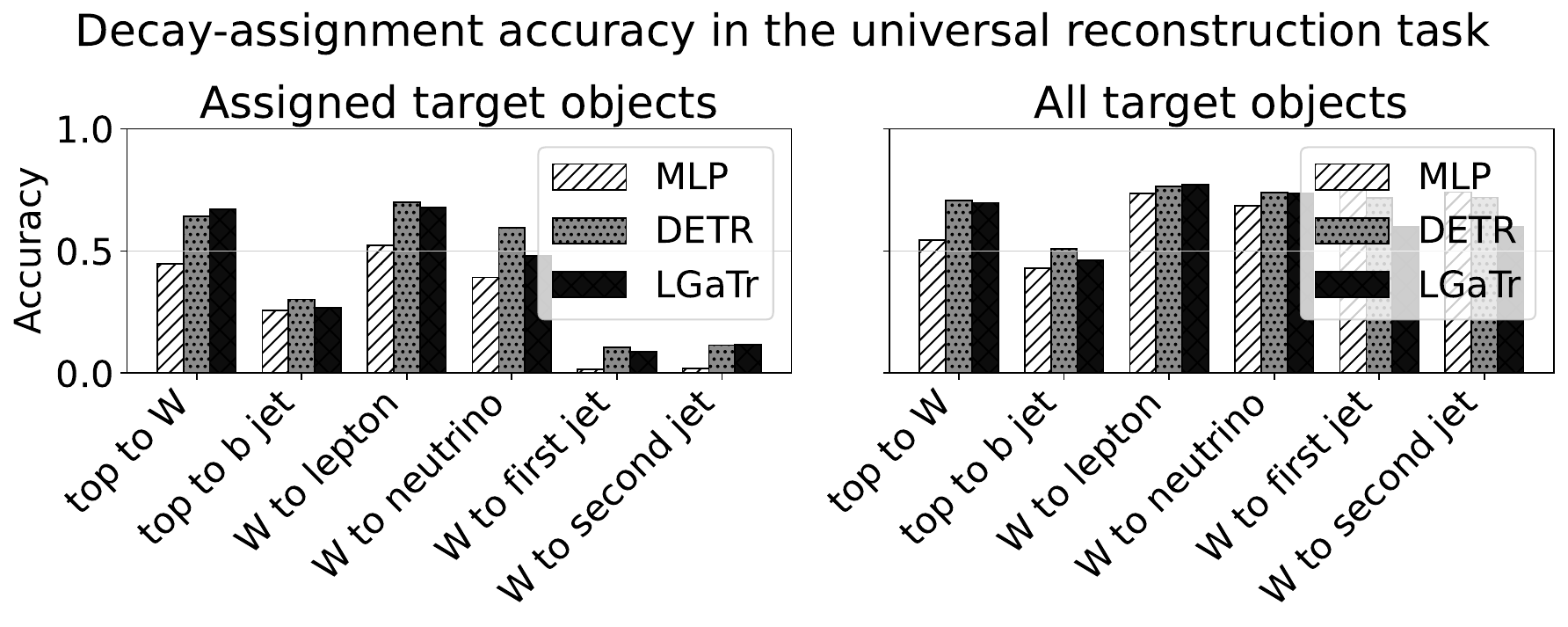}\par\medskip
    \includegraphics[width=0.55\linewidth]{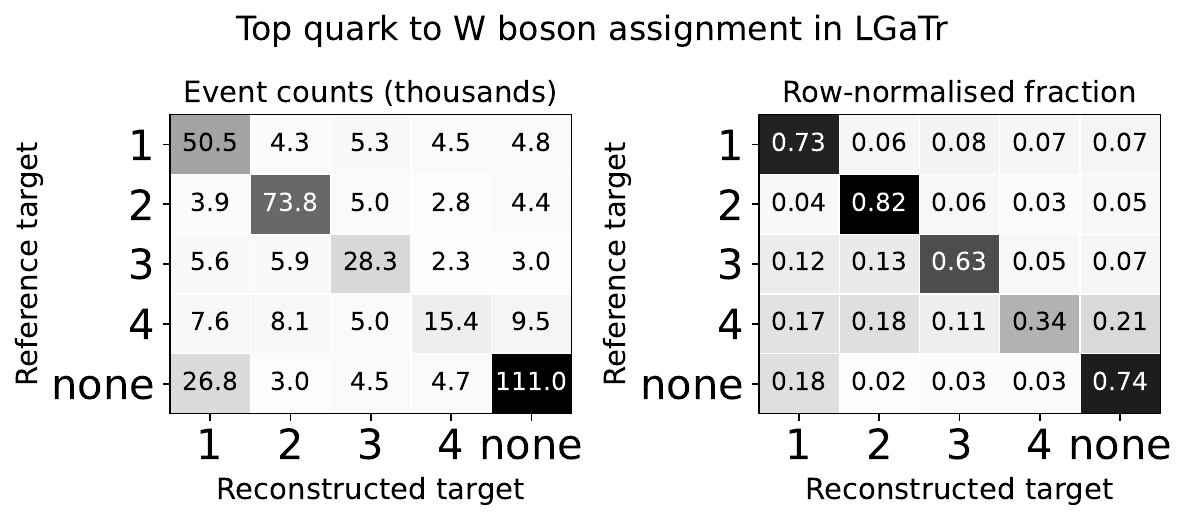}
    \caption{LGaTr classification and decay-assignment results in the universal
    event-reconstruction test sample.  The upper row gives LGaTr top quark
    classification, and the following row gives lepton classification.  Their
    matrices give reference classes (rows) and reconstructed classes (columns),
    and their charts give class accuracy; higher values are better.  The assignment charts
    compare accuracy (vertical axis) for six relations on the horizontal axis:
    top quark to \(W\) boson, top quark to b jet, and \(W\) boson to its lepton,
    neutrino, first jet, and second jet.  The left chart retains only target
    slots with a physical reference child, excluding no-match; the right chart
    includes all valid target slots, including no-match.  Diagonal, dotted, and
    crossed hatching denote MLP, DETR, and LGaTr, respectively.  The lower
    matrices show the LGaTr assignment of top quarks to \(W\) bosons.  Reference
    (rows) and reconstructed (columns) top quark target indices are labelled
    1--4 or ``none''.  The left matrix reports counts in thousands; the right
    gives fractions after each reference-target row is normalized to unity.}
    \label{fig:top-cls}
    \label{fig:lepton-cls}
    \label{fig:univ-net-assignment}
    \label{fig:top-w-assignment}
\end{figure}

\begin{figure}[H]
    \centering
    \includegraphics[width=0.9\linewidth]{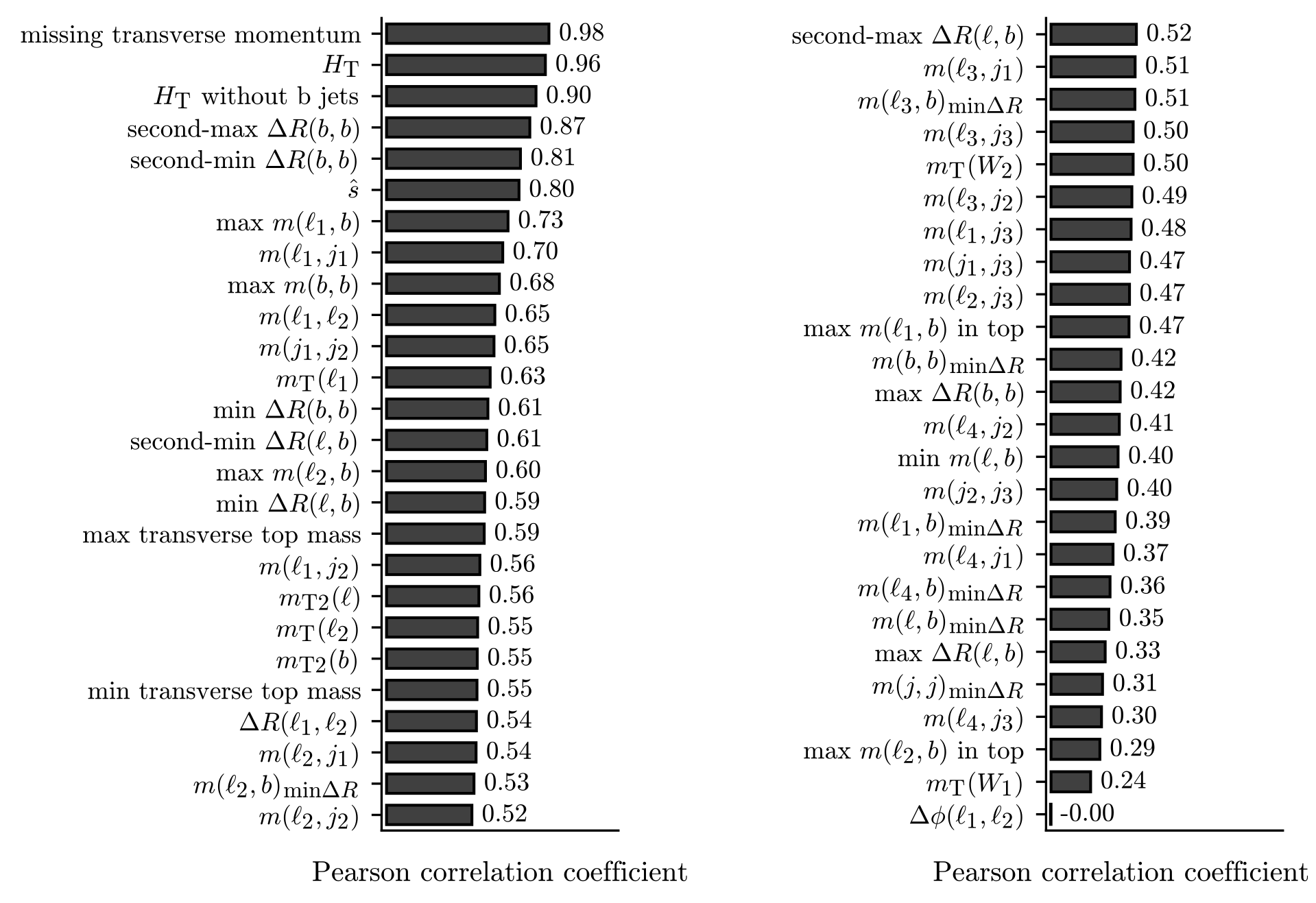}
    \caption{Pearson correlation coefficients between LGaTr-reconstructed and
    target high-level observables in the universal event reconstruction sample.
    Larger correlations indicate stronger reconstruction.}
    \label{fig:hlr}
\end{figure}

From the performance evaluation plots for HLR in Fig.~\ref{fig:hlr}, assignment in Fig.~\ref{fig:univ-net-assignment}, classification in Fig.~\ref{fig:top-cls} and regression in Fig.~\ref{fig:top-reg} one can infer that the LGaTr performs on par with DETR-like architecture and significantly outperforms MLP in assignment - the most difficult subtask. Having said that, there are several places where LGaTr requires additional improvements to either input handling or the architecture itself:
\begin{itemize}
    \item $W\to j$ asssignment is poor for all networks - this is caused by a significant class imbalance - all samples are focused on leptonic decays, thus most of the $W\to j$ target slots are empty and models do not have enough signal data to learn passable assignment. The same issue arises with third/fourth top decays.
    \item Some important HLR variables are poorly reconstructed, accompanied by the wrong $\phi$ angle predictions in regression outputs. Is is assumed that $\phi$ is the main culprit, and some additional changes in LGaTr are needed to enforce flat distribution of this variable.
\end{itemize}

\subsection{Prospects for improvement}

Larger samples of hadronic \(W\) boson decays, high-multiplicity top quark
categories, and dedicated task weights for occupied assignments can increase
the contribution of sparse targets to the loss.  An angle-periodic readout and
task-aware high-level regression weights are possible future refinements which should contribute to better HLR and regression quality.

\FloatBarrier

\section{Conclusions}
\label{sec:conclusions}

A common Lorentz-covariant language for typed event representations and for
operations on their multivector components is provided by geometric algebra.  Different tasks with
different input definitions are addressed by the $t\bar{t}/tW$ separationm and the flavour-changing neutral-current
single top quark production study. GA-based networks perform on par or better than baseline MLPs, which are trained on hand-crafted features.

The comparison is extended to a shared multiprocess, multi-task setting with detector tokens and learned parton queries by the universal event-reconstruction configuration. In this task LGaTr outperforms baseline MLP on all tasks, but falls short in reconstruction of high-level variables and 4-momenta regression compared to DETR-like architecture.

Further improvements include more hadronic \(W\) boson decays,
high-multiplicity top quark categories, task-aware weights for occupied
assignments and high-level regression, and an angle-periodic readout, which should contribute to better HLR and regression quality.

\section*{FUNDING}
This study was conducted within the scientific program of the National Center for Physics and Mathematics, section \#5 «Particle Physics and Cosmology». Stage 2026-2027

\renewcommand{\bibfont}{\scriptsize}

\end{document}